# The hazard ratio is interpretable as an odds or a probability under the assumption of proportional hazards

**David M. Thompson**
**J.E. Reid**

Three statistical studies, all published between 2004 and 2008 but without referring to one another, assert a useful equivalence involving the hazard ratio, a parameter estimated for time to event data by the frequently used proportional hazards model.

Stated generally (if somewhat tediously), when the assumption of proportional hazards is justified, the hazard ratio *is equivalent to the odds* that a randomly chosen person from the group whose hazard is represented in the ratio's numerator will experience the event of interest *before* a randomly chosen person from the group represented by the ratio's denominator.

## *The three studies*

The three studies use the equivalence to facilitate communication about research results that are quantified using a hazard ratio. They carefully avoid confusing the hazard ratio with a risk ratio or relative risk.

The first of the three to be published (Spruance, Reid, Grace, Samore, 2004) point out that, when hazards are proportional, "the hazard ratio is equivalent to the odds that an individual in the group with the higher hazard reaches the endpoint first" (p. 2790).

The next year, Henderson and Keiding (2005) ask the reader to "consider two patients, one with low risk and one with high risk. They refer to relative risks (h) rather than hazard ratios, and maintain (p.705) that "the probability that the high risk patient will live longer [experience death, the event of interest, later] than the low risk patient" is $1/(1+h)$. Therefore, "the rate ratio h is equal to the odds that the high risk patient dies before the low risk patient."

Finally, Moser and McCann (2008) demonstrate that the hazard ratio estimated by a Cox proportional hazard model is equivalent to a parameter $\theta$. In contrast to the two earlier studies, they define the parameter as a probability, not as an odds. However, they point out (p. 250) that "there exists a one-to-one algebraic equivalence between the hazard ratio $\Delta$ and the probability $\theta$" and their Equation (1) verifies that the hazard ratio $\Delta$ is an odds.

Thus, Moser and McCann describe $\theta$ as the probability (and equivalently the odds) "that a patient given one treatment will have an event earlier than if the same patient were given a different treatment." Moser and McCann's study is the only one of the three to use this "counterfactual" language, and it may represent an overreach of logic. However, overlooking that distinction, Moser and McCann equate the hazard ratio with the probability or odds that an event occurs earlier in individuals from one group than in the other.

*Using the equivalence to communicate about research results*

Of the three studies, the one that appeared first (by Spruance, Reid and colleagues) offers the most direct advice to physicians and others on how to use the equivalence to communicate research results to patients.

> "Thus, in a clinical trial examining time to disease resolution, [the hazard ratio] represents the odds that a treated patient will resolve symptoms before a control patient. Stated another way, for any randomly selected pair of patients, one from the treatment group and one from the control group, the hazard ratio is the odds that the time to healing is less in the patient from the treatment group than in the patient from the control group."

To communicate with those who find probabilities more intuitive than odds, they observe (p. 2790) that "the probability of healing first can easily be derived from the odds of healing first, which is the probability of healing first divided by the probability of not healing first: hazard ratio (HR) = odds = P/(1-P); P= HR/(1+ HR). A hazard ratio of 2 therefore corresponds to a 67% chance of the treated patient's healing first, and a hazard ratio of 3 corresponds to a 75% chance of healing first".

To the question "what are the chances I will do better on this new drug compared to no treatment," they recommend that the physician reply: "The odds are roughly 2:1 (the probability is 66%) that you will have an episode of shorter duration than someone who did not take the drug" (p. 2791).

*A derivation for the equivalence*

Spruance's co-author, J.E. Reid, suspected the equivalence in 2000, while performing simulations, when she noted that samples with the same hazard ratio also had the same values for Harrell's c-statistic. In a time-to-event analysis, the c statistic denotes the proportion of pairs of individuals, randomly drawn from the sample such that both experienced the outcome (neither were censored), for which the statistical model correctly predicts the member of the pair who experienced the outcome first. The utility of c-indexes is explored in detail by Combescure and colleagues (2014).

Spruance's article did not include the group's derivation for the equivalence. The derivation below follows their scenario concerning healing times.

Following Spruance, Reid and colleagues, consider the event of interest to be healing or disease resolution.

Let Y be the random time to event in the treatment group, and regard this group as representing the estimated hazard ratio's numerator.

Let X be the random time to event in the control group, and regard this group as representing the estimated hazard ratio's denominator.

The probability that a randomly chosen person in the treatment group experiences the event *after* a randomly chosen person in the control group is p(Y>X).

Define the following functions *for the control group,*

> Let h(t) be the hazard function
> Let S(t) be the survival function
> Let f(t) be the probability density function, such that f(t)=h(t)S(t)

Let $\lambda$ be the hazard ratio. Assuming the groups' hazards are proportional, then the survival function *for the treatment group* is $S(t)^\lambda$.

The joint probability of Y and X, such that Y>X, is equal to the product of the conditional probability p(t>X) and the marginal probability f(t), which is the probability density function for the control group:

$$P(Y>X) = \int_0^\infty P(t>X) f(t)dt$$

$$= \int_0^\infty P(t>X) h(t)S(t)\, dt$$

Recognizing that p(t>X) relates to the survival function of the treatment group, which, assuming hazards are proportional, is equal to $S(t)^\lambda$:

$$P(Y>X) = \int_0^\infty S(t)^\lambda h(t) S(t)\, dt$$

A u substitution that defines:
> u=S(t)
> du/dt = dS(t)/dt = -f(t) = -h(t)S(t)
> du=-h(t)S(t)dt

yields:

$$p(Y>X) = \int_0^\infty h(t)S(t)S(t)^\lambda dt = \int_0^\infty -u^\lambda\, du = \left[-\frac{1}{\lambda+1} S(t)^{\lambda+1}\right]_{t=0}^{t=\infty} = -\frac{1}{\lambda+1}(0-1) = \frac{1}{\lambda+1}$$

The probability p(Y>X) that a randomly chosen person in the treatment group experiences the event (healing) *after* a randomly chosen person in the control group is $1/(\lambda+1)$.

Therefore, p(Y<X), the probability that a randomly chosen person in the treatment group experiences the event (healing) *before* a randomly chosen person in the control group, is

> 1 - p(Y>X) = $\lambda/(\lambda+1)$,

The probability $\lambda/(\lambda+1)$ is recognizable as being equivalent to an odds of $\lambda$. The hazard ratio $\lambda$ therefore represents the odds that Y<X, that is, the odds that a randomly chosen member of the treatment group experiences the event (healing, in this scenario) before a randomly chosen member of the control group.